\documentclass[aps,prd,twocolumn,fleqn,superscriptaddress,preprintnumbers]{revtex4}

\usepackage{graphicx,color}
\usepackage{amsmath,amssymb,amsfonts}

\newcommand{\be}{\begin{equation}}
\newcommand{\ee}{\end{equation}}

\newcommand{\bea}{\begin{eqnarray}}
\newcommand{\eea}{\end{eqnarray}}

\newcommand{\ba}{\begin{eqnarray}}
\newcommand{\ea}{\end{eqnarray}}
\renewcommand{\(}{\left(}
\renewcommand{\)}{\right)}

\newcommand{\lk}{\left[}
\newcommand{\rk}{\right]}

\newcommand{\w}{\hat{\omega}}

\def\be{\begin{equation}}
\def\ee{\end{equation}}
\def\beq{\begin{eqnarray}}
\def\eeq{\end{eqnarray}}

\def\g{\gamma}

\begin{document}

\input amssym.def
\input amssym.tex

\title{Thermalization at intermediate coupling}

\preprint{BI-TP 2012/39, TUW-12-23}

\author{Dominik Steineder}
\affiliation{Institute of Theoretical Physics, Technical University of Vienna,\\ Wiedner Hauptstr.~8-10, A-1040 Vienna, Austria}
\author{Stefan A. Stricker}
\affiliation{Institute of Theoretical Physics, Technical University of Vienna,\\ Wiedner Hauptstr.~8-10, A-1040 Vienna, Austria}
\author{Aleksi Vuorinen}
\affiliation{Faculty of Physics, University of Bielefeld,
D-33615 Bielefeld, Germany}

\begin{abstract}
We use the AdS/CFT conjecture to investigate the thermalization of large-$N_c$ ${\mathcal N}=4$ Super Yang-Mills plasma in the limit of large but finite 't Hooft coupling. On the gravity side, we supplement the type IIB supergravity action by the full set of ${\mathcal O}(\alpha'^3)$ operators, which enables us to derive ${\mathcal O}(\lambda^{-3/2})$ corrections to the emission spectrum of prompt photons in one model of holographic thermalization. Decreasing the coupling strength from the $\lambda=\infty$ limit, we observe a qualitative change in the way the photon spectral density approaches its thermal limit as a function of the photon energy. We interpret this behavior as a sign of the thermalization pattern of the plasma shifting from top/down towards bottom/up.
\end{abstract}

\maketitle

%%%%%%%%%%%%%%%%%%%%%%%%%%%%%%%%%%%%%%%%%%%
{\em Introduction.}
%%%%%%%%%%%%%%%%%%%%%%%%%%%%%%%%%%%%%%%%%%%
Figuring out the precise way, in which quark gluon plasma created in a heavy ion collision approaches local thermal equilibrium prior to its eventual freeze-out and hadronization, is a question of utmost importance for a successful description of the system. In the limit of weak coupling, the thermalization pattern is known to be of the bottom/up type, with the soft excitations reaching the thermal limit before the hard ones \cite{Baier:2000sb}; recently, it has been found that the driving force behind this behavior is related to plasma instabilities \cite{Kurkela:2011ti,Kurkela:2011ub}. In contrast, studies of thermalization in the strongly coupled limit of conformal field theories, utilizing the gauge/gravity duality \cite{Maldacena:1997re,Witten98,GKP98}, have pointed towards top/down type thermalization \cite{Balasubramanian:2011ur,CaronHuot:2011dr,Chesler:2011ds,Galante:2012pv,Hubeny:2010ry,Erdmenger:2012xu} (for an overview of holographic calculations in the heavy ion context, see also \cite{CasalderreySolana:2011us})
. This indicates the probable existence of a transition between the two behaviors at intermediate coupling. 

The range of 't Hooft couplings relevant for a heavy ion collision, $\lambda\equiv g^2N_c\sim 20$, falls clearly outside the realm of weak coupling techniques (see e.g.~\cite{Chesler:2006gr} for a discussion of this issue). Starting from the opposite end, $\lambda=\infty$, one might however wonder, if a consistent inclusion of strong coupling corrections might allow one to access this region. In the case of large-$N_c$ ${\mathcal N}=4$ Super-Yang-Mills (SYM) theory, these corrections are available through the inclusion of the full set of ${\mathcal O}(\alpha'^3)$ type IIB string theory operators in the supergravity action \cite{Paulos:2008tn, Myers:2008yi}, resulting in corrections to different physical quantities typically proportional to the parameter $\gamma \equiv \frac{1}{8} \zeta(3)\lambda^{-3/2}$ \cite{Gubser:1998nz, Pawelczyk:1998pb, Buchel:2004di, Hassanain:2011fn}. These calculations are often conceptually straightforward but technically very involved, one prominent example being the recent 
determination of the (thermal) photon production rate beyond its well-known $\lambda=\infty$ limit \cite{CaronHuot:2006te} in refs.~\cite{Hassanain:2011ce,Hassanain:2012uj}. In these articles, the authors first took advantage of their earlier results \cite{Hassanain:2009xw} to derive ${\mathcal O}(\lambda^{-3/2})$ corrections to the Schr\"odinger equation of a bulk U(1) vector field, and then proceeded to solve this equation to obtain the photon spectral density on the field theory side.

In an altogether different generalization of ref.~\cite{CaronHuot:2006te}, two of the present authors studied photon and dilepton production in an out-of-equilibrium ${\mathcal N}=4$ SYM plasma \cite{Baier:2012tc,Baier:2012ax}, working within a model of holographic thermalization that involves the gravitational collapse of a thin spherical shell in AdS$_5$ space \cite{Danielsson:1999fa,Lin:2008rw}. This model is homogeneous in three-space and allows an analytic treatment of photon and dilepton spectral densities in the limits of high frequencies and late times. In these two cases, characterized by the slow relative motion of the shell, one may namely approach the dynamics of the system in a quasistatic approximation, where the shell is treated as a static object when formulating the boundary conditions of different bulk fields. This presents an important simplification in the calculations, and in particular allows one to verify the validity of the fluctuation dissipation theorem, necessary to connect the 
field theory spectral densities to the corresponding (photon/dilepton) production rates.

In the paper at hand, our plan is to combine the approaches of refs.~\cite{Hassanain:2011ce,Hassanain:2012uj,Baier:2012ax} to study the production of prompt photons in a thermalizing ${\mathcal N}=4$ SYM plasma beyond the usual $\lambda=\infty$ limit. In particular, we aim to extract the relative deviation of the photon spectral density from its thermal limit in a specific out-of-equilibrium state, and to examine the dependence of this quantity on photon energy at different couplings. As we will discuss in the following sections, we argue that this information can be used to study the pattern, with which plasma constituents of different energy approach local thermal equilibrium.

%%%%%%%%%%%%%%%%%%%%%%%%%%%%%%%
{\em The setup.}
%%%%%%%%%%%%%%%%%%%%%%%%%%%%%%%%%%%%%
We work within a model of holographic thermalization that involves the gravitational collapse of a thin shell of matter in AdS$_5$ space. The radial coordinate of the shell (in a coordinate system where the boundary is located at $r=\infty$) is denoted by $r_s$. It satisfies at all (finite) times the relation $r_s>r_h$, where $r_h$ is the Schwarzschild radius of the eventual black hole, corresponding to the temperature of the field theory in its final thermal state. The dynamics of the shell, i.e.~the dependence of $r_s$ on the field theory time $t$, can in principle be solved from the corresponding equation of motion; cf.~the discussion in \cite{Baier:2012tc}. For the purposes of the present paper, we however leave this question aside, as we will in any case work within the quasistatic approximation, where the thermalization process is most conveniently parametrized by the values of $r_s/r_h$.

Inside the shell, the metric of the spacetime is that of pure AdS$_5$, while outside it is given by an AdS black hole solution; the effects of being out of thermal equilibrium then become manifested through the changing boundary conditions of different bulk fields at the location of the shell. The model was initially introduced in ref.~\cite{Danielsson:1999fa} and further analyzed e.g.~in \cite{Erdmenger:2012xu,Lin:2008rw} (see also \cite{Wu:2012ri} for a discussion of how the collapsing shell can be realized by turning on a scalar source on the boundary). Our notation and conventions follow those of ref.~\cite{Baier:2012tc}, to which we refer the interested reader for more details.

In order to study photon production, we next add to the ${\mathcal N}=4$ SYM theory a gauge field coupled to a conserved current corresponding to one of the U(1) subgroups of the SU(4) R-symmetry of the theory. As discussed in \cite{CaronHuot:2006te}, this is most conveniently performed in a way, in which all ${\mathcal N}=4$ scalars and fermions obtain equal charges, leading to an anomaly free theory. To leading order in the electromagnetic coupling $\alpha$ and to all orders in $\lambda$, the photon production rate per unit volume is then given by the formula
\be\label{rate}
k^0 \frac{d\Gamma_{\gamma}}{d^3k} = \frac{1}{4\pi k}\frac{d\Gamma_{\gamma}}{dk_0} =
\frac{\alpha}{4 \pi^2} \eta^{\mu\nu} \Pi_{\mu \nu}^{<} (k^0 = k) ~,
\ee
where $\Pi_{\mu \nu}^{<}$ is the electromagnetic current Wightman function. In thermal equilibrium as well as in the quasistatic limit of the falling shell setup (see the discussion in appendix B of \cite{Baier:2012tc}), the fluctuation dissipation theorem further allows one to relate $\Pi_{\mu \nu}^{<}$ to the transverse photon spectral density $\chi_{\mu\nu}$,
\be
\eta^{\mu\nu} \Pi_{\mu \nu}^{<} (k^0 = k) \, = \,n_B(k^0) \chi_{\mu}^{\mu}(k_0)\,\equiv\, n_B(k^0) \chi(k_0)  ~,\label{wight}
\ee
with $n_B(k_0)=(e^{k_0/T}-1)^{-1}$. This implies that it suffices to consider the retarded correlator of the electromagnetic (in practice, R symmetry) current in the SYM theory, which simplifies the dual gravity calculation in a dramatic way.

As we want to study photon production to the next-to-leading order in a small curvature expansion, we must include $\mathcal{O}(\alpha'^3)$ corrections to the type IIB supergravity action. This results in \cite{Paulos:2008tn, Myers:2008yi}
\ba
S_{IIB}&=&\frac{1}{2\kappa_{10}^2}\int d^{10}x\sqrt{-G}\bigg(R_{10}-\frac{1}{2}(\partial \phi)^2\nonumber \\
&&-\frac{F_5^2}{4\cdot 5!}+\g e^{-\frac{3}{2}\phi}(C+\mathcal{T})^4\bigg)\, , \label{action}
\ea
where $R_{10}$ is the 10-dimensional Ricci scalar, $\phi$ the dilaton, and $F_5$ the five-form field strength, and where we have set the curvature radius of AdS$_5$ space to unity. In the ${\mathcal O}(\gamma)$ terms, $C$ denotes the 10-dimensional Weyl tensor, while $\mathcal{T}$ stands for a complicated tensor built from the five-form and defined e.g.~in eq.~(5) of \cite{Hassanain:2011ce}. 

The $\gamma$-corrected AdS black hole metric derived from the above action has the form \cite{Gubser:1998nz,Pawelczyk:1998pb,Paulos:2008tn}
\ba
ds^2 &=& \frac{r_h^2}{u} \, \left(-f(u) \,
K^2(u) \, dt^2 + d\vec{x}^2\right) \nonumber \\
&&+ \frac{1}{4 u^2
f(u)} \, P^2(u) \, du^2 + L^2(u) \, d\Omega_5^2\, ,
\ea
where $f(u)= 1-u^2$ and $u\equiv r_h^2/r^2$ is a dimensionless coordinate, in which the boundary of the AdS space is located at $u=0$. The different functions appearing here are given by
\ba
\!\! K(u) &=& e^{\gamma \, [a(u) + 4b(u)]}\,, \,\, P(u) = e^{\gamma \,
b(u)}\,, \,\, L(u) = e^{\gamma \, c(u)}\, , \nonumber \\
a(u) &=& -\frac{1625}{8} \, u^2 - 175 \, u^4 + \frac{10005}{16} \,
u^6 \, , \nonumber \\
b(u) &=& \frac{325}{8} \, u^2 + \frac{1075}{32} \, u^4
- \frac{4835}{32} \, u^6 \, , \nonumber \\
c(u) &=& \frac{15}{32} \, (1+u^2) \, u^4 \,,
\ea
while the $\gamma$-corrected relation between $r_h$ and the field theory temperature reads $r_h=\pi T/(1+\frac{265\g}{16})$.

%%%%%%%%%%%%%%%%%%%%%%%%%%%%%%%%%%%%%%%%
{\em The calculation.}
%%%%%%%%%%%%%%%%%%%%%%%%%%%%%%%%%%%%%%%%
To determine the required R current correlator including all necessary $\gamma$-corrections is in principle a very laborious exercise. Fortunately, we can use here the results of refs.~\cite{Hassanain:2011ce,Hassanain:2009xw,Hassanain:2012at}, in which the authors consider a vector field living in the above background metric. Specializing to the momentum space component of a transverse electric field, $E_\text{T}(u,\omega)$, with $\omega$ the corresponding frequency, they arrive at a Schr\"odinger-type action
\ba
S&=&-\frac{N_c^2r_h^2}{16\pi^2}\int_k \int \!{\rm d}u\, \bigg[\frac{1}{2}\Psi{\mathcal L}\Psi +\partial_u \Phi \bigg]\, ,
\ea
where the boundary term reads $\Phi(u)=\Psi'(u)\Psi(u)$. The relation between $\Psi$ and $E_\text{T}$ can be found from \cite{Hassanain:2012at} and has the form $\Psi(u) \equiv \Sigma(u) E_\text{T}(u)$, $\Sigma(u)^{-1}\equiv1/\sqrt{f(u)}+\gamma p(u)$, with the function $p(u)$ given by
\ba
p(u)&=&\frac{u^2(11700 - u^2 (343897 + 37760 \w^2 u - 87539 u^2))}{288\sqrt{f(u)}} \, , \nonumber \\
\hat{\omega}&\equiv& \omega/(2\pi T)\, .
\ea
With these definitions, the equation of motion for the field $\Psi(u)$ takes the simple form $\Psi''(u)-V(u)\Psi(u)=0$, where the potential $V(u)$ reads
\ba
\!\!\!\!\!\!\!\!V(u)&=&-\frac{1}{f(u)^2}\Bigg(1+\hat{\omega}^2 u -\frac{\gamma}{144}f(u) \Big[-11700\\
&&+2098482u^2-4752055u^3+1838319u^6\nonumber \\
&&+\hat{\omega}^2 u\(\-16470+245442u^2+1011173u^4\)\Big]\Bigg)\, . \nonumber
\ea

Working in the collapsing shell model, we need to solve the above equation for $\Psi$ outside the shell, i.e.~for $u<u_s$. Inside the shell, $u>u_s$, neither the (pure AdS) metric nor the equation of motion for $\Psi$ is altered by $\gamma$-corrections \cite{deHaro:2003zd, Banks:1998nr}, and thus the solution to the latter can be read off e.g.~from ref.~\cite{Baier:2012tc}. Here, one only needs to recall the relation between the frequencies measured inside and outside the shell,
\be
\w_-=\frac{\w_+}{\sqrt{f_m}}\equiv\,\frac{\w}{\sqrt{f_m}} , \quad f_m\,\equiv\, f(u_s)K^2(u_s)\, ,
\ee
where the $-$ subscript refers to the inside and + to the outside. The two solutions furthermore need to satisfy a set of junction conditions at the shell \cite{Baier:2012tc}, 
\ba\label{mc}
\!\!\Psi_{-}(u_s)&=&\sqrt{f_m}\Psi_+(u_s)/\Sigma(u_s)\, , \\
\!\!\Psi'_-(u_s)&=&f_m \partial_u\big(\Psi_+(u)/\Sigma(u)\big)\big|_{u=u_s}\nonumber \, ,
\ea
which result from demanding continuity of the physical electric field and where the $u$-derivative on the left hand side of the second relation does not operate on the parameter $u_s$ in $\hat{\omega}_-$.

%%%%%%%%%%%%%%%%%%%%%%%%%%%%%%%%%%%%%%%
\begin{figure}
\centering
\includegraphics[width=8.0cm]{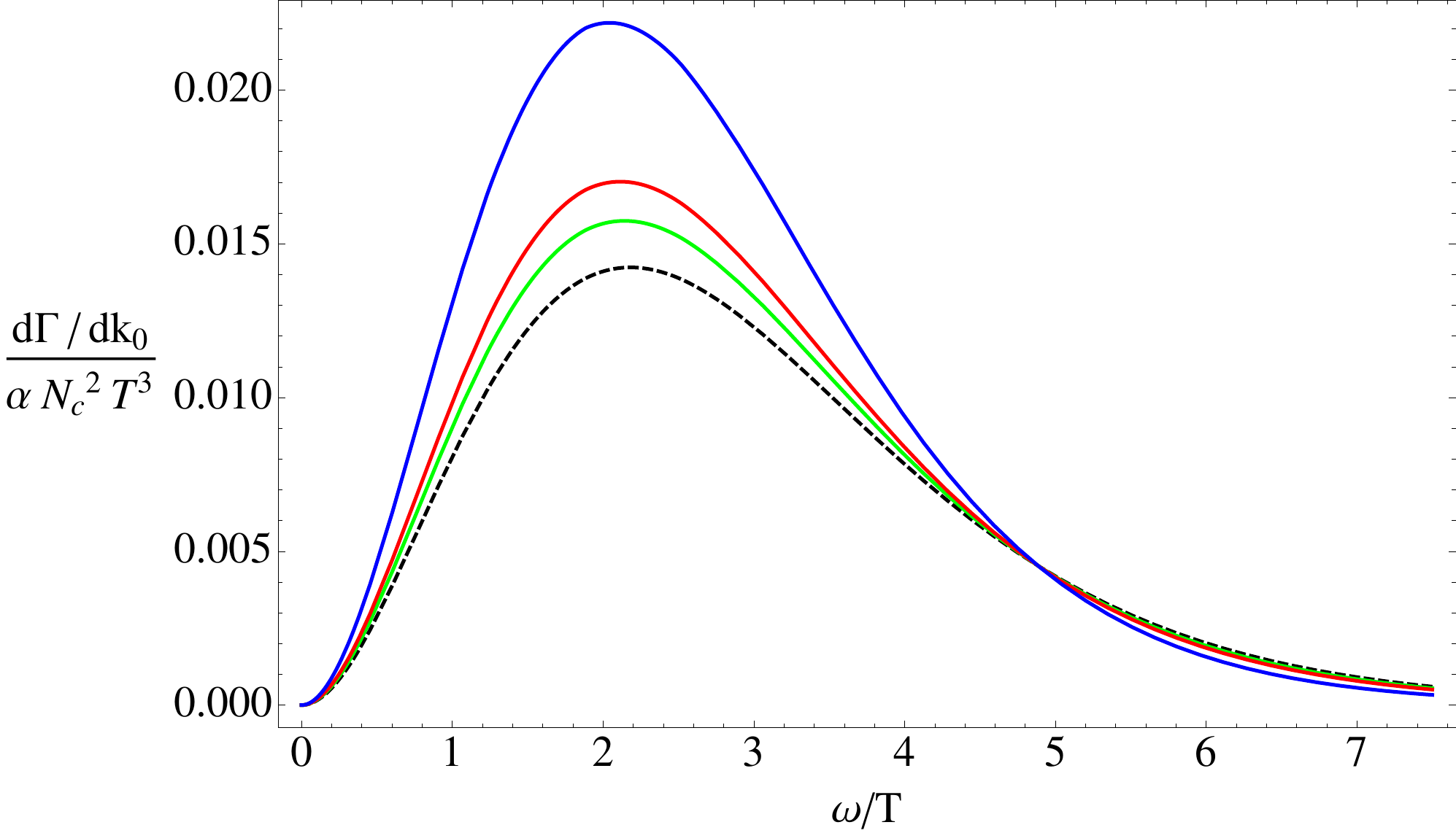}
\caption {The photoemission rate $d\Gamma_\gamma/k_0$, normalized by $\alpha N_c^2 T^3$ and evaluated for $r_s/r_h=1.01$ and $\lambda=\infty,\,120,\,80,\,40$ (from bottom to top at the peak value). The qualitative form of the curves stays unaltered for all small and moderate values of $r_s/r_h$.}
\label{photonrate}
\end{figure}
%%%%%%%%%%%%%%%%%%%%%%%%%%%%%%%%%%%%%%%

Following the above steps, we first obtain for the inside solution
\be
\Psi_{-}(u)=\Psi_{-}^0(u)+\g \Psi_{-}^1(u)\, ,
\ee 
where $\Psi_{-}^0$ can be read off from eq.~(10) of \cite{Baier:2012ax} (noting that inside the shell $\Psi(u)=E_\text{T}(u)$) and where the correction term $\Psi_{-}^1$ is available through a simple expansion of the modified Bessel functions. For the outside solution, $\Psi_+(u)$, we on the other hand write
\be
\Psi_{+}=c_-(\Psi_\text{in}^0+\g \Psi_\text{in}^1)+c_+(\Psi_\text{out}^0+\g \Psi_\text{out}^1)\, ,
\ee
where the subscripts `in' and `out' refer to infalling and outgoing field modes at $u=1$. The zeroth order solutions can again be read off from ref.~\cite{Baier:2012ax} (after a rescaling by $\sqrt{f(u)}$), while the correction terms are obtained through a numerical solution of the equation of motion discussed in the previous section, subject to the relevant boundary conditions. Matching this outside solution to the inside one via eq.~(\ref{mc}) then provides us with a $\gamma$-corrected result for the ratio $c_-/c_+$, through which we have determined the behavior of $\Psi(u)$ in the entire AdS space.

%%%%%%%%%%%%%%%%%%%%%%%%%%%%%%%%%%%%%%%%
{\em Results.}
%%%%%%%%%%%%%%%%%%%%%%%%%%%%%%%%%%%%%%%%
To obtain the photon spectral density $\chi$ on the field theory side, the above results simply need to be inserted into the relation \cite{SonStarinets}
\be
\chi(\w)=\frac{N_c^2 T^2}{2}\(1-\frac{265}{8}\g\)\mathrm{Im}\lk \frac{\Psi'_{+}}{\Psi_{+}}\rk \Bigg|_{u=0}\, ,
\ee
which we expand to linear order in $\g$. From here, we easily obtain both the relative deviation of the spectral density from its equilibrium limit,
\be
R(\w)=\frac{\chi(\w)-\chi_\text{th}(\w)}{\chi_\text{th}(\w)},
\ee
as well as the corresponding photon production rate,
\begin{equation}
\frac{d\Gamma_\gamma}{dk_0}=\frac{\alpha k}{\pi}n_B(k_0) \chi(k_0)\Big|_{k_0=k=2\pi T\w}\, .
\end{equation}
Our results for these quantities are to be compared on one hand to the $\lambda=\infty$ off-equilibrium results of ref.~\cite{Baier:2012ax} and on the other hand to the ${\mathcal O}(\alpha'^3)$ corrected equilibrium calculation of ref.~\cite{Hassanain:2011ce}.

%%%%%%%%%%%%%%%%%%%%%%%%%%%%%%%%%%%%%%%
\begin{figure}
\centering
\includegraphics[width=8.0cm]{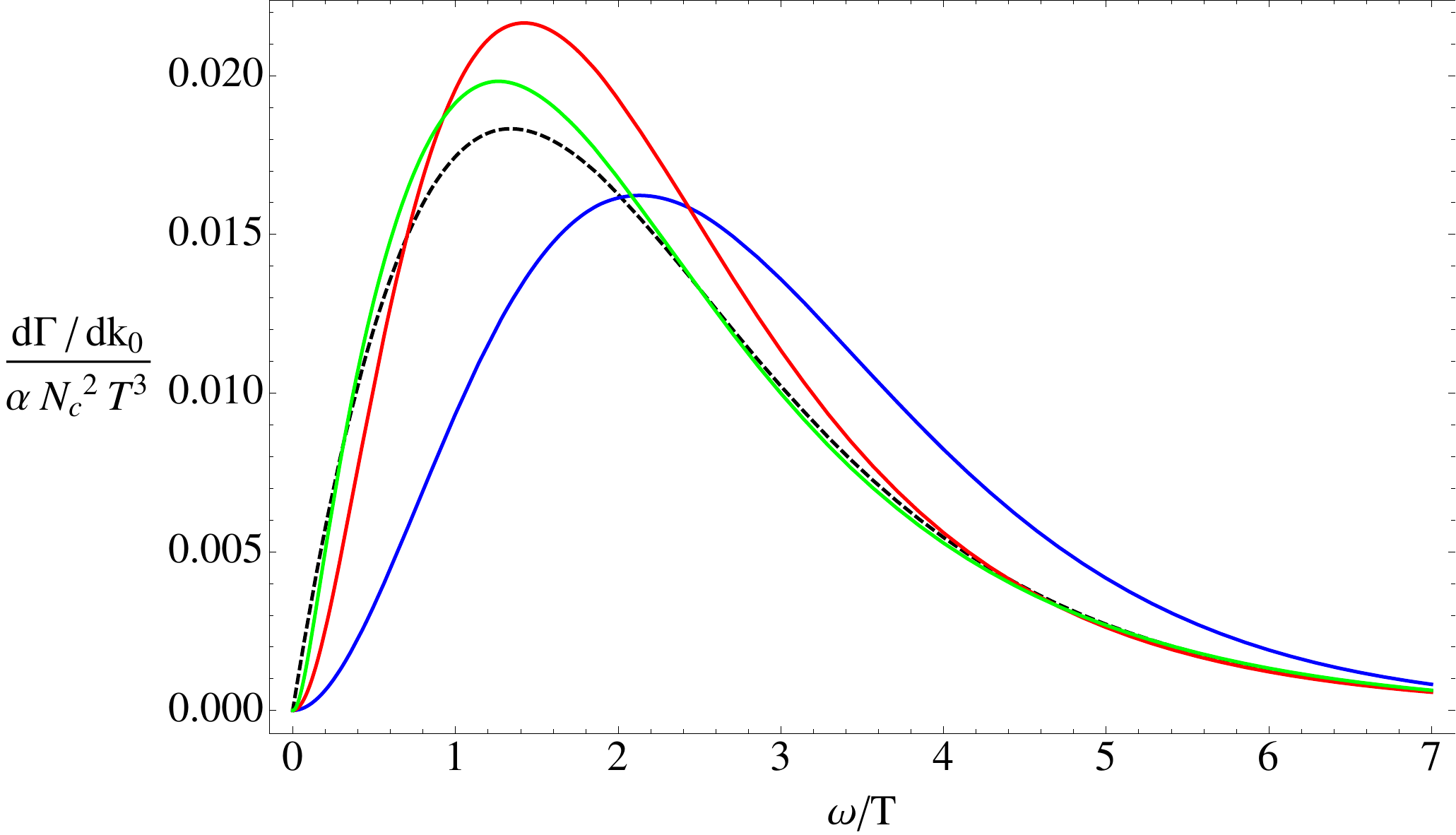}
\caption {The photoemission rate $d\Gamma_\gamma/k_0$, normalized by $\alpha N_c^2 T^3$ and shown for $\lambda=100$ and $r_s/r_h=1.1,\, 1.01,\, 1.001,\, 1$ (from bottom to top at very small frequencies).}
\label{photonrate2}
\end{figure}
%%%%%%%%%%%%%%%%%%%%%%%%%%%%%%%%%%%%%%%

In fig.~\ref{photonrate}, we first display the photoemission rate for different values of $\lambda$ (chosen such that the strong coupling expansion is still applicable), with the shell always residing at $r_s/r_h=1.01$. The pattern we observe is very similar to that described in the thermal case in ref.~\cite{Hassanain:2011ce}: Decreasing the 't Hooft coupling from $\lambda=\infty$, the peak of the spectrum increases and moves towards smaller $\omega$. At the same time, fixing the value of the coupling to $\lambda=100$ and varying $r_s/r_h$ is seen to largely reproduce the qualitative findings reported in ref.~\cite{Baier:2012ax} for $\lambda=\infty$; see fig.~\ref{photonrate2} above. The main difference between the behavior of our curves and those plotted in fig.~3 of \cite{Baier:2012ax} is the faster shifting of the peak of the finite-$\lambda$ spectral density towards larger $\omega$, when $r_s/r_h$ is increased. This has the effect of suppressing the value of the peak due to the appearance of the Bose-Einstein distribution in eq.~(\ref{rate}), best seen in the blue $r_s/r_h=1.1$ curve in fig.~\ref{photonrate2}.

%%%%%%%%%%%%%%%%%%%%%%%%%%%%%%%%%%%%%%%
\begin{figure*}
\centering
\includegraphics[width=7.8cm]{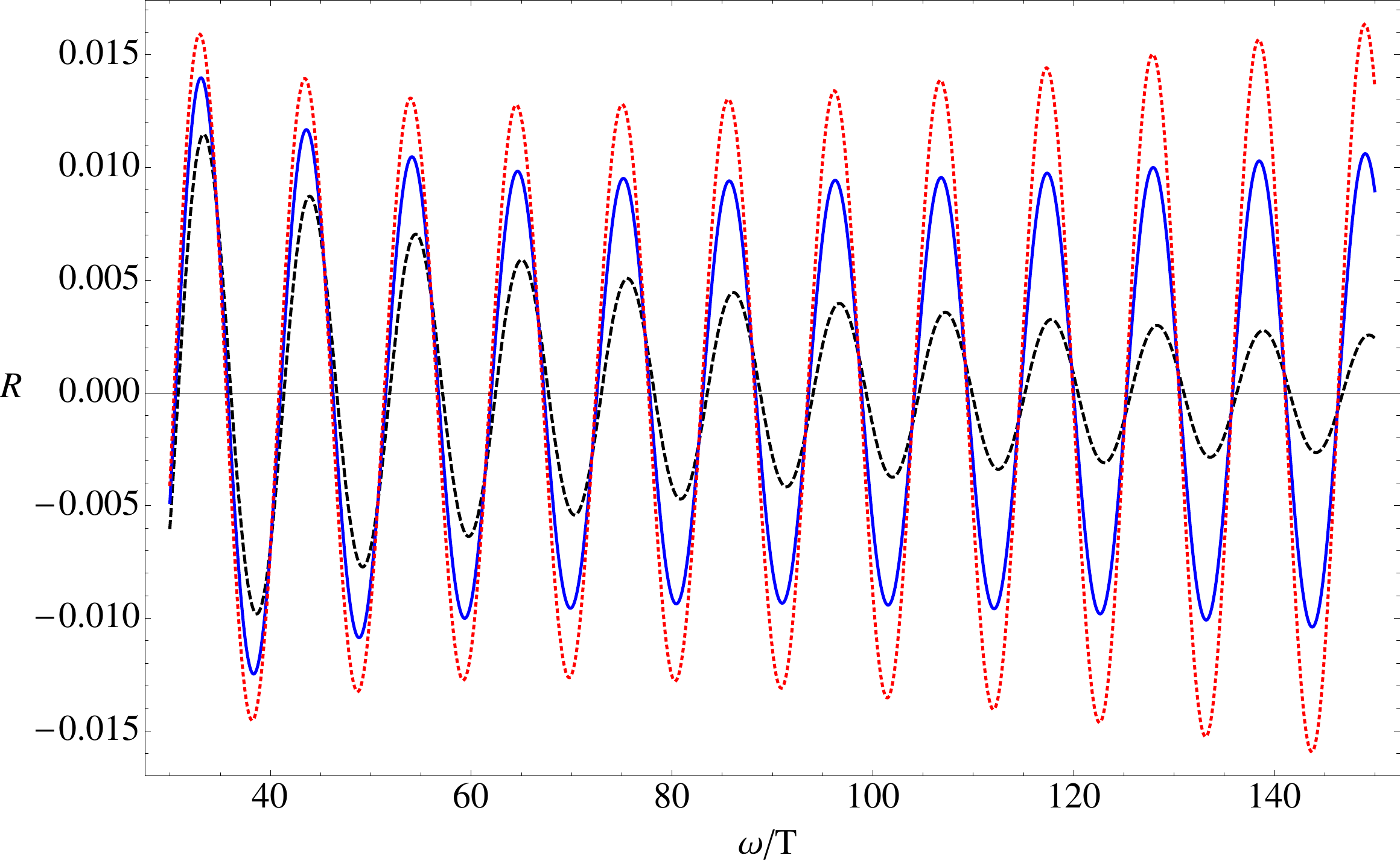}$\;\;\;\;\;\;\;\;$\includegraphics[width=7.8cm]{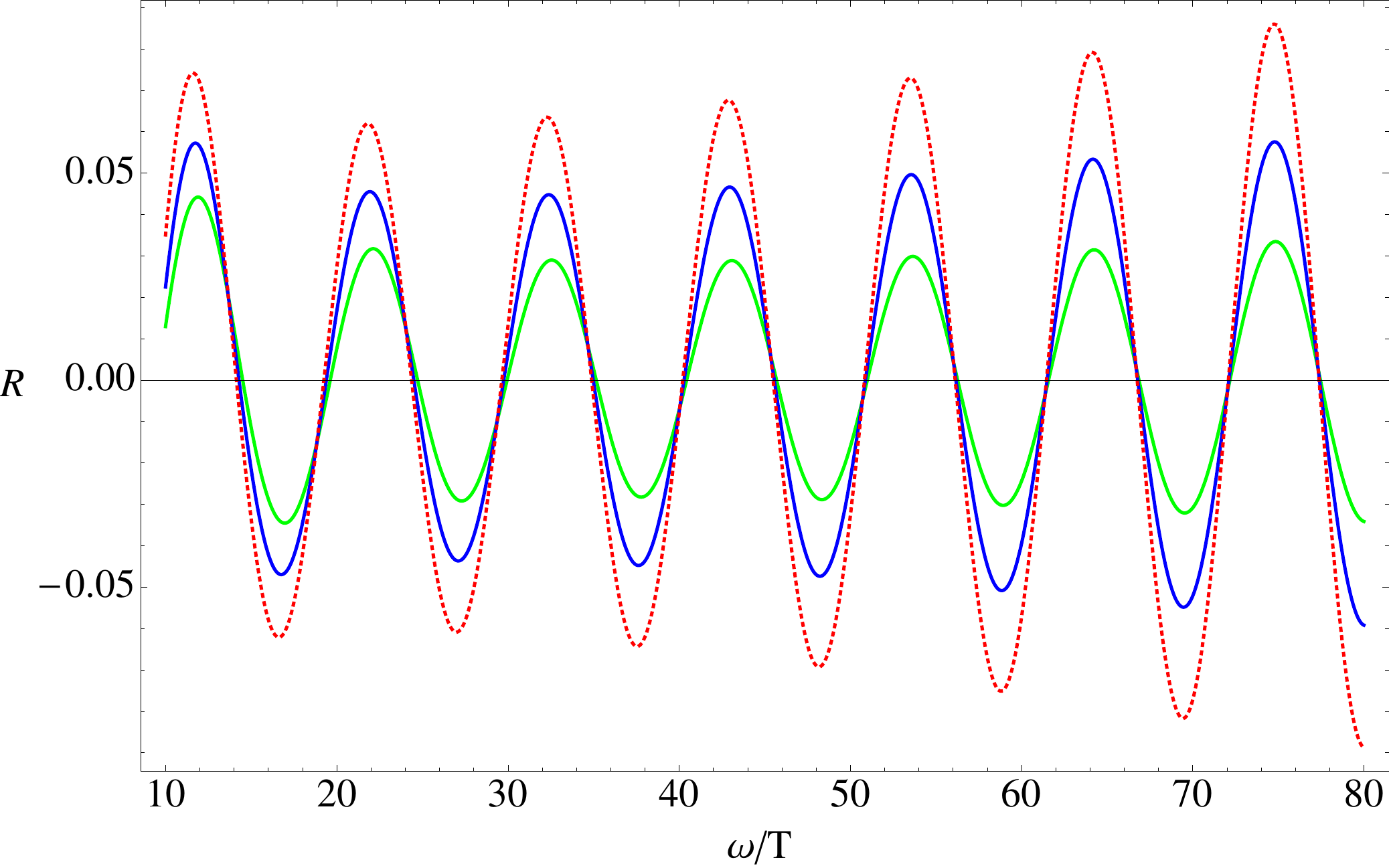}
\caption {The relative deviation $R$, plotted for $r_s/r_h=1.01$ and $\lambda=\infty,\,500,\,350$ (left) and $\lambda=150,\, 100,\, 75$ (right), with the amplitudes of the curves increasing with decreasing coupling. While in the $\lambda=\infty$ case the amplitude of the oscillations gets damped at large $\omega$, for all finite values of $\lambda$ it first decreases but ultimately starts growing linearly with $\omega$.}
\label{R}
\end{figure*}
%%%%%%%%%%%%%%%%%%%%%%%%%%%%%%%%%%%%%%%

In fig.~\ref{R}, we display our perhaps most prominent finding, the behavior of the relative deviation of the spectral density from its thermal limit, $R(\w)$, for $r_s/r_h=1.01$ and $\lambda=\infty$, 500, and 350 (left), as well as 150, 100 and 75 (right). The quantity is observed to exhibit oscillations similar to those discussed already in \cite{Baier:2012ax}, which increase in frequency and decrease in amplitude as the shell approaches the horizon. Interestingly, the behavior of the fluctuations as functions of $\omega$ seems to depend on the value of $\lambda$ rather strongly: As soon as one leaves the strict $\lambda=\infty$ limit, the asymptotic large-$\omega$ behavior of the amplitude of $R$ changes from a $1/\omega$ suppression to a linear enhancement. For a given value of $\lambda$, the amplitude of $R$ is in fact observed to have a minimum at some $\omega=\omega_\text{min}(\lambda)$, signifying the frequency, for which the spectral density is closest to its thermal limit. This is in direct contrast with the usual geometric picture of strong coupling top/down thermalization, where one expects the hardest modes to always thermalize first. A numerical study further shows that within the validity of the strong coupling expansion, the function $\omega_\text{min}(\lambda)$ always obeys a power law behavior $\omega_\text{min}(\lambda)/T \approx \alpha \lambda^\beta$, where the positive constants $\alpha$ and $\beta$ depend on the value of $r_s/r_h$ only very mildly. For $r_s/r_h=1.01$, we obtain $\alpha\approx 1.25$, $\beta\approx 0.69$.

Although it is good to recall that the photon frequency $\omega$ does not directly correspond to the energy of the plasma constituents, it is tempting to speculate that the above behavior of $R$ might be indicative of the thermalization pattern of the plasma changing from top/down towards bottom/up as the value of $\lambda$ is decreased. After all, it is at the very least clear that the highly energetic photons with $\omega \gg T$ cannot have been emitted by the soft excitations of the plasma. Furthermore, one should note that our observation regarding the shift in the asymptotic behavior of $R$ is extremely robust. First, it is seen at all non-infinite values of $\lambda$, including a regime where the strong coupling expansion is guaranteed to converge. And second, going beyond the limit of a static shell would result in corrections proportional to $1/(\tau\omega)$, where $\tau$ is the characteristic time scale related to the motion of the shell. As discussed in section 4 of \cite{Baier:2012tc}, these contributions are expected to be strongly suppressed for large enough values of $\omega$, such as those considered in fig.~\ref{R} above.

%%%%%%%%%%%%%%%%%%%%%%%%%%%%%%%%%%%%%%%%
{\em Conclusions.}
%%%%%%%%%%%%%%%%%%%%%%%%%%%%%%%%%%%%%%%%
In the paper at hand, we have used the AdS/CFT conjecture to study the behavior of an out-of-equilibrium large-$N_c$ ${\mathcal N}=4$ SYM plasma at large, yet finite 't Hooft coupling. The calculation was carried out within one particular model of holographic thermalization, where the approach of the plasma towards thermal equilibrium is modeled via the gravitational collapse of a thin shell of matter in AdS$_5$ space. In this setup, we were able to derive the leading ${\mathcal O}(1/\lambda^{3/2})$ corrections to the photon production rate and the corresponding spectral density. Our most interesting finding, displayed in fig.~\ref{R}, was interpreted to reflect a change in the thermalization pattern of the plasma from top/down towards bottom/up as $\lambda$ is decreased. In particular, our results indicate that for values of $\lambda$ of relevance to real life heavy ion collisions, one may already be quite far from a strict top/down behavior.

Although our results offer only indirect evidence of the pattern, with which strongly coupled SYM plasma thermalizes, we feel that they constitute an interesting starting point for further study. An immediate generalization of the present calculation should be a similar determination of energy momentum tensor correlators within the collapsing shell model, providing direct information on the behavior of the plasma constituents themselves. On top of this, it would be intriguing to extend the present calculation beyond the quasistatic approximation, as well as to other, more realistic models of holographic thermalization. These studies would be crucial to assess the universality of our findings, as well as the conclusions to be drawn. If it turned out that the observed change in the thermalization pattern of a non-Abelian large-$N_c$ plasma with decreasing $\lambda$ was a generic prediction of holographic calculations, it would clearly highlight the importance of including strong coupling corrections to the present studies. At the same time, such a result would be highly encouraging for the future prospects of modeling thermalization in heavy ion collisions using holography.

%%%%%%%%%%%%%%%%%%%%%%%%%%%%%%%%%%%%%%%%%%%%
{\em Acknowledgments.}
%%%%%%%%%%%%%%%%%%%%%%%%%%%%%%%%%%%%%%%%
We are grateful to Martin Schvellinger for helpful comments and for providing us access to an early version of \cite{Hassanain:2012at}. We also thank Janne Alanen, Rolf Baier, Alex Buchel, Keijo Kajantie, Ville Ker\"anen, Esko Keski-Vakkuri, Aleksi Kurkela, Anton Rebhan, Olli Taanila and Bin Wu for useful discussions. D.S.~was supported by the Austrian Science Foundation FWF, project no.~P22114, S.S.~and A.V.~by the Sofja Kovalevskaja program of the Alexander von Humboldt Foundation, and S.S.~by the FWF START project Y435-N16.

%%%%%%%%%%%%%%%%%%%%%%%%%%%%%%%%%%%%%%%%%%%%%%%%%%%%%%

\end{document}